\documentclass[aps,twocolumn]{revtex4}
\usepackage{epsf}

\begin{document}
 
%\title{Particle-hole mixing of preformed quasiparticle pairs\\ 
%       in the pseudogap state of superconductors}

\title{Unconventional particle-hole mixing in the systems \\
       with strong superconducting fluctuations}

\author{T.\ Doma\'nski}
\affiliation{
           Institute of Physics, M.\ Curie Sk\l odowska University, 
           20-031 Lublin, Poland 
}
\date{\today}

\begin{abstract}

Development of the STM and ARPES spectroscopies enabled to reach 
the resolution level sufficient for detecting the particle-hole 
entanglement in superconducting materials. On a quantitative level 
one can characterize such entanglement in terms of the, so called,  
Bogoliubov angle  
%introduced by P.W.\ Anderson 
which 
%becomes experimentally accessible. This angle 
determines to what extent the particles and holes constitute 
the spatially or momentum resolved excitation spectra.
In classical superconductors, where the phase transition is related 
to formation of the Cooper pairs almost simultaneously accompanied 
by onset of their long-range phase coherence, the Bogoliubov angle 
is slanted all the way up to the critical temperature $T_{c}$. In 
the high temperature superconductors and in superfluid ultracold 
fermion atoms near the Feshbach resonance the situation is different 
because of the preformed pairs which exist above $T_{c}$ albeit 
loosing coherence due to the strong quantum fluctuations. We 
discuss a generic temperature dependence of the Bogoliubov angle 
in such pseudogap state indicating a novel, non-BCS behavior. For 
quantitative analysis we use a two-component model describing 
the pairs coexisting with single fermions and study their mutual 
feedback effects by the selfconsistent procedure originating from 
the renormalization group approach.

%Motivated by a novel ability of STM spectroscopy to measure the Bogolubov 
%angle we propose to use this technique for identifying precursor 
%effects above the transition temperature $T_{c}$ of underdoped cuprate 
%superconductors. Within a phenomenological two-component model we show 
%that {\em the Bogolubov angle can emerge already in the normal state} 
%upon approaching $T_{c}$ from above. Its measurements might clear up 
%what part of the pseudogap region in the phase diagram of cuprates 
%corresponds to the Bogolubov-like quasiparticles and, ultimately 
%whether $T^{*}$ is related to superconducting fluctuations.
\end{abstract}

\maketitle

\section{Introduction} 

Such vastly distinct systems as the classical and/or high $T_{c}$ cuprate superconductors, 
the ultracold superfluid fermion atoms as well as certain cosmological (superfluid 
neutron stars) and even subatomic objects (odd-odd nuclei) reveal signatures 
of ideally coherent pairs consisting of particles from a vicinity of the Fermi 
surface. Obviously, what differs one case from another is an underlying mechanism 
and energy scale engaged in the pairing. They all however share the universal 
feature related to the effective Bogoliubov quasiparticles representing 
a superposition of the fermion particles and their absence. This emerging 
particle-hole (p-h) mixing \cite{Fujita-07} has a purely quantum nature 
(imposed by the structure of the BCS wave function) which to some extent 
resembles the corpuscular-wave dualism. One of its spectacular manifestations 
is the mechanism of subgap Andreev reflection where an incident fermion-particle 
can convert into the pair with a simultaneous reflection of the fermion-hole 
what is indeed observed experimentally in superconductors \cite{Deutcher-05}, 
for the relativistic-like particles \cite{Beenakker-08} and in quantum dots 
attached to superconducting electrodes \cite{Domanski-Andreev,QD}.

In the recent papers A.\ Balatsky and coworkers have emphasized that p-h 
mixing can be quantitatively probed by the present-day STM \cite{Fujita-07} 
and ARPES spectroscopies \cite{angle_ARPES}. These techniques are capable 
to determine either the spatially \cite{Fisher-07} or momentum resolved 
\cite{Damascelli-04} single particle excitation spectra of superconductors. 
In principle also the simultaneous {\bf k}- and {\bf r}-space measurements 
are feasible by means of the Fourier transformed quasiparticle interference 
imaging \cite{McElroy-08}. Roughly speaking, the p-h mixing manifests itself 
in the single particle spectra by appearance of two peaks around the Fermi 
level separated by twice the (pseudo)gap and whose spectral weigths 
yield the information on particle/hole contributions to the Bogoliubov 
quasiparticles. 

Usually for conventional superconductors these contributions are given 
by the BCS coefficients $u_{\bf k}^{2}$ and $v_{\bf k}^{2}=1-
u_{\bf k}^{2}$, so it is convenient to define the Bogoliubov 
angle \cite{remark}
\begin{eqnarray}
\theta_{\bf k} = \frac{\pi}{2} - 2 \; \mbox{arctan} \left( 
\frac{ |u_{\bf k}|}{| v_{\bf k}| }  \right)
\label{BA}
\end{eqnarray}
as a measure of the particle-hole mixing. Its magnitude can vary 
between $-\pi/2$ and $\pi/2$ depending on a momentum and indirectly 
on temperature. $\theta_{\bf k}$ has a particularly clear interpretation
in the pseudospin representation $\hat{s}_{{\bf k},z}\!=\!\frac{1}
{2}\left(1\!-\hat{c}_{{\bf k}\uparrow}^{\dagger}\hat{c}_{{\bf k}
\uparrow}\!-\!\hat{c}_{-{\bf k}\downarrow}^{\dagger}\hat{c}_{-{\bf k}
\downarrow}\right)$, $\hat{s}_{{\bf k},x(y)}\!=\! \frac{1}{2(i)}
\left(\hat{c}_{{\bf k}\uparrow}^{\dagger}\hat{c}_{{-\bf k}\downarrow}
^{\dagger}\!+(-) \hat{c}_{{-\bf k}\downarrow}\hat{c}_{{\bf k}\uparrow}
\right)$ introduced by P.W.\ Anderson \cite{Anderson-58}, where it denotes 
an azimuthal angle of the vector $\langle \hat{\bf s}_{\bf k}\rangle$. 
Restricting to the part of Hilbert space where $\langle \hat{c}_{{\bf k}
\uparrow}^{\dagger}\hat{c}_{{\bf k}\uparrow}\rangle\!=\!\langle \hat{c}
_{-{\bf k}\downarrow}^{\dagger}\hat{c}_{-{\bf k}\downarrow}\rangle$ 
the pseudospin eventually points down (up) when effective quasiparticles 
are represented by particles (holes). The upper and bottom panels of 
figure \ref{Fig1} illustrate such behavior well known for the normal 
and superconducting states \cite{Anderson-58}. 

In general, pseudospins obey the non-trivial dynamics governed by 
the Bloch-type equations of motion \cite{Anderson-58}. This aspect
has a particular importance in the context of ultracold atoms where   
traversing through the Feshbach resonance can lead to the soliton-like 
solutions \cite{dynamics}. On the other hand, in the highly 
inhomogeneous cuprate superconductors with pairing on a local 
(interactomic) distance both the excitation spectrum \cite{Fisher-07} 
and the Bogoliubov angle are strongly varying in space. 
Such issue has been already explored within the Bogoliubov de 
Gennes approach and results were confronted with the available 
STM data \cite{Fujita-07}. 

Since the Bogoliubov angle (\ref{BA}) is sensitive to existence 
of the paired fermions one may ask if any signatures of the p-h 
mixing would be able to appear above $T_{c}$. ARPES studies 
\cite{Matsui-03} confirm that the superconducting state of cuprates 
obeys roughly the usual BCS behavior but there is still no firm 
agreement on the nature of pseudogap state and its relation to 
superconductivity \cite{last_reviews}. Nevertheless, various 
experimental data \cite{pairs_above_Tc,ARPES_2008} seem to indicate 
that preformed fermions' pairs are present already in the normal state 
(at least in the underdoped samples) at temperatures up to dozen 
Kelvin above $T_{c}$. Transition temperature might there correspond 
to the onset of long-range phase coherence \cite{fluct_Tc}. Another 
evidence of the preexisting pairs above $T_{c}$ is known for 
the ultracold atoms of Li$^{6}$ and K$^{40}$. Near the Feshbach 
resonance the weakly bound boson molecules are scattered into 
the Cooper-like pairs and such unitary limit is in a crossover 
between the BCS and BEC regimes beeing influenced by strong 
quantum fluctuations \cite{cold_atoms}.

Our purpose here is to explore the impact of preformed pairs 
on the Bogoliubov angle in the pseudogap state. In particular, 
we address the question whether p-h mixing can at all show up 
above $T_{c}$ and if so, then how it would manifest itself. For 
the considerations we use a phenomenological two-component model 
\cite{Ranninger-85} where itinerant fermions and their paired 
counterparts are introduced without referring to any specific 
microscopic mechanism. From the selfconsistent treatment of 
interactions between the paired and single fermions we find 
the evidence of particle-hole mixing signified by $|\theta
_{\bf k}| \!\neq\! \pi/2$. Furthermore, lack of the phase 
coherence above $T_{c}$  leads to a discontinuity of $\theta
_{\bf k}$ at ${\bf k}_{F}$. We will show that in the pseudogap 
state the Bogoliubov angle behaves in a manner which partly 
resembles the normal and partly the superconducting phases 
(see figure \ref{Fig1}).

In the next section we briefly introduce the model and discuss 
its main properties. Methodological details are presented in 
section III and the essential part on the p-h mixing for the 
pseudogap state is described in section IV. We finally summarize 
our results and point out some related unresolved problems.

\section {Phenomenological model}

For modelling the pseudogap state we use the following Hamiltonian
\cite{Ranninger-85}
\begin{widetext}
\begin{eqnarray}
\hat{H} & = & \sum_{{\bf k},\sigma} \left( \varepsilon_{\bf k}\!-\! 
\mu \right) \hat{c}_{{\bf k}\sigma}^{\dagger} \hat{c}_{{\bf k}\sigma} 
+ \sum_{\bf q} \left( E_{\bf q}\!-\!2\mu \right) 
\hat{b}_{\bf q}^{\dagger} \hat{b}_{\bf q} \label{BF} 
%\\ & + & 
\;+\;\frac{1}{\sqrt{N}} \; \sum_{{\bf k},{\bf q}} \left(  
g_{{\bf k},{\bf q}} \hat{b}_{\bf q}^{\dagger} \hat{c}_{{\bf q}-{\bf k}
\downarrow}\hat{c}_{{\bf k}\uparrow} + g_{{\bf k},{\bf q}}^{*}
\hat{c}_{{\bf k}\uparrow}^{\dagger} \hat{c}_{{\bf q}-{\bf k}\downarrow}
^{\dagger}\hat{b}_{\bf q} \right) \;,
%\nonumber
\end{eqnarray}
%
%    ----------   Figure 1   ----------    %
\begin{figure}
\epsfxsize=13cm
\centerline{\epsffile{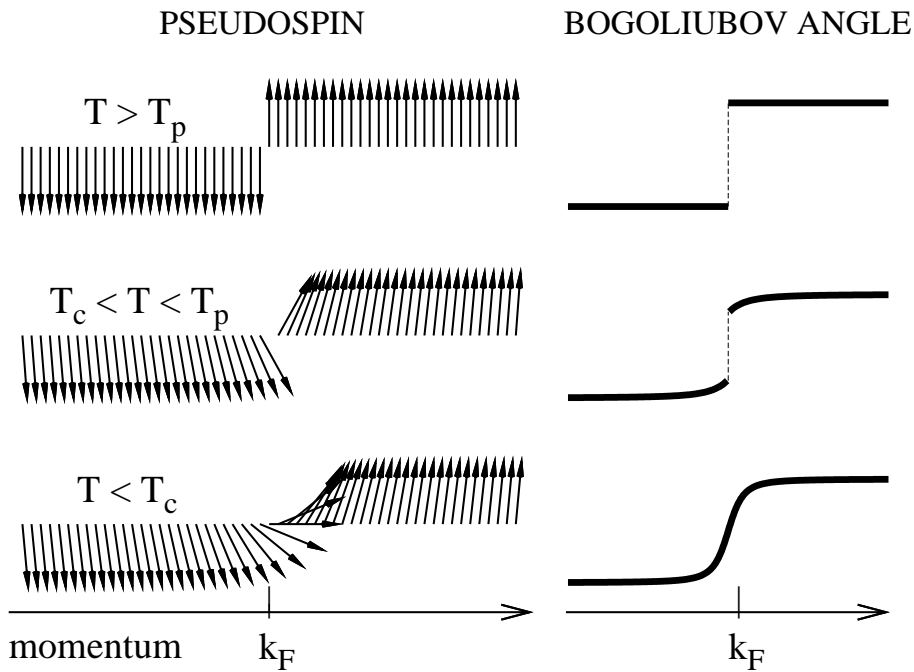}}
\caption{Variation of the Anderson's pseudospin (the left h.s.\ column) 
and the Bogoliubov angle $\theta_{\bf k}$ (the right h.s.\ column) against 
momentum in the normal, pseudogap and superconducting states. 
Notice that particle-hole mixing is present in the superconducting 
and pseudogap states, however above $T_{c}$ the Bogoliubov angle 
becomes discontinuous at ${\bf k}_{F}$.}
\label{Fig1}
\end{figure}
%     -------------------------------     %     
\end{widetext}
where operators $\hat{c}_{{\bf k}\sigma}^{(\dagger)}$ refer to annihilation
(creation) of single fermions with the energy $\varepsilon_{\bf k}$ and 
$\hat{b}_{\bf q}^{(\dagger)}$ correspond to the local pairs of energy
$E_{\bf q}$. Potential of the interaction between the single and paired 
fermions is denoted by $g_{{\bf k},{\bf q}}$. For simplicity, we shall 
assume that concentration of pairs per lattice site is small enough so 
that $\hat{b}_{\bf q}^{(\dagger)}$ obey the usual bosonic commutation 
relations (we neglect the hard-core effect).

This model (\ref{BF}) has been invented \cite{Ranninger-85} and explored 
by J.\ Ranninger with coworkers \cite{Ranninger-all} and independently 
by T.D.\ Lee et al \cite{TD_Lee} as well as some other groups. Starting 
from various microscopic models several authors \cite{various_authors} 
have also concluded that the relevant physics of strongly correlated 
cuprates is well captured by the fermion and boson degrees of freedom 
expressed by the Hamiltonian (\ref{BF}). Moreover, such model well 
describes the ultracold fermion atoms interacting 
with the Feshbach resonance \cite{cold_atoms,Levin-05}. 

In the simplest mean-field approach one can linearize the interaction 
term so that the decoupled boson and fermion parts become exactly 
solvable \cite{Ranninger-85}. The resulting spectrum of fermions has 
then BCS structure $A^{MF}({\bf k},\omega)=u^{2}_{\bf k} \delta 
( \omega - E_{\bf k} )+ v^{2}_{\bf k} \delta ( \omega + E_{\bf k} )$ 
with the usual quasiparticle energy $E_{\bf k}=\sqrt{ ( 
\varepsilon_{\bf k}\!-\!\mu )^{2}+\Delta_{\bf k}^{2}}$ and 
coherence factors $u^{2}_{\bf k},v^{2}_{\bf k}=\frac{1}{2}
[1\!\pm\!( \varepsilon_{\bf k} \!-\!\mu )/E_{\bf k}]$ which 
lead to the standard Bogoliubov angle. Energy gap of the single 
particle excitation spectrum is effectively given by $\Delta_{\bf k}
\!=\!g_{{\bf k},{\bf 0}} \sqrt{ \langle n^{B}_{\bf 0}\rangle}$. 
This means that fermions undergo transition to the superconducting 
state if and only if the Bose-Einstein condensation of bosons 
takes place \cite{Ranninger-85}. Actually, the latter property 
is valid exactly \cite{Kostyrko-96} without limitations to any 
approximation.

The mean-field treatment does not take into account the quantum 
fluctuations whose efficiency increases upon approaching $T_{c}$ 
and above of it. In the next section we present the method which 
enables a selfconsistent study of the boson-fermion feedback effects. 
In particular, we will analyze the remnants of superconducting 
correlations above $T_{c}$ and study their effect on the Bogoliubov 
angle.

\section{The procedure}

For studying the model (\ref{BF}) we use the selfconsistent,
non-perturbative procedure based on a canonical transformation 
$\hat{H} \!\longrightarrow\! e^{\hat{S}(l)}\hat{H}e^{-\hat{S}(l)}$ 
with a continuous formal parameter $l$ \cite{Wegner-94}. The main idea 
is to eliminate the interaction part $g_{{\bf k},{\bf q}}$ through 
a sequence of infinitesimal steps $l \rightarrow l + \delta l$. 
Proceeding along the lines of the Renormalization Group (RG) technique 
one starts from renormalizing the high energy sector and subsequently 
turns to the low energy sector (by latter we mean the fermion states 
close to $\mu$ and boson states near $2\mu$). We briefly describe 
some technicalities in order to clarify how the particle and hole 
spectral contributions can be evaluated within this procedure.

Practically we start by setting $\hat{H}(l)\!\equiv\!e^{\hat{S}(l)}
\hat{H}e^{-\hat{S}(l)}$, where $\hat{H}(0)$ corresponds to the initial 
Hamiltonian, and then construct the flow equation $\partial_{l}\hat{H}(l)
\!=\! [ \hat{\eta}(l),\hat{H}(l)]$ with the generating operator 
$\hat{\eta}(l) \equiv \partial_{l}\hat{S}(l)$. Following the original 
proposal of Wegner \cite{Wegner-94} we choose $\hat{\eta}(l)\!=\!
[\hat{H}_0(l),\hat{H}_{int}(l)]$, where $\hat{H}_{0}(l)$ denotes the 
total kinetic energy of fermions and bosons whereas $\hat{H}_{int}(l)$ 
stands for their interaction. From a straightforward algebra we obtain 
$\hat{\eta}(l)\!=\!- \frac{1}{\sqrt{N}} \sum_{{\bf k},{\bf q}}\alpha
_{{\bf k},{\bf q}}(l) \left(  b_{\bf q}^{\dagger}c_{{\bf q}
-{\bf k} \downarrow} c_{{\bf k} \uparrow} - \mbox{h.c.}\right)$ with 
$\alpha_{{\bf k},{\bf q}}(l)=\left( \varepsilon_{\bf k}(l)+
\varepsilon_{{\bf q}-{\bf k}}(l)-E_{\bf q}(l)\right) g_{{\bf k},
{\bf q}}(l)$. One can prove analytically \cite{Domanski-01} that 
such antihermitean operator $\hat{\eta}(l)$ indeed guaranties 
an asymptotic disappearance of the boson-fermion coupling 
$\lim_{l\rightarrow\infty} \;g_{{\bf k},{\bf q}} (l)\!=\!0$.

%    ----------   Figure 2   ----------    %
\begin{figure}
\epsfxsize=10cm
\centerline{\epsffile{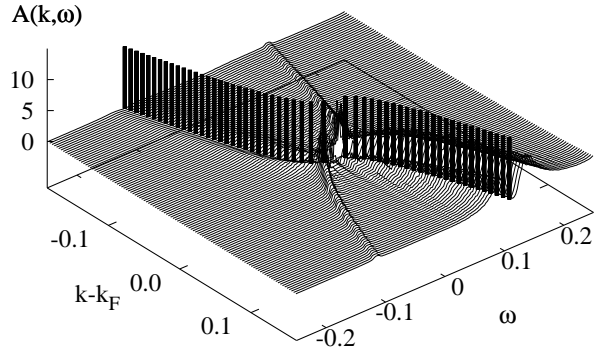}}
\caption{The single particle excitation spectrum of fermions
in the pseudogap regime. Besides the long-lived quasiparticle
at $\omega\!=\!\tilde{\varepsilon}_{\bf k}\!-\!\mu$ there
emerges its mirror reflection corresponding to the damped
Bogoliubov shadow branch whose presence has been confirmed 
by the recent ARPES measurements \cite{ARPES_2008}. 
Both branches are separated by the pseudogap which vanishes 
at $T_{p}\!>\!T_{c}$.}
\label{Fig2}
\end{figure}
%     -------------------------------     %     

Applying this scheme to the boson-fermion Hamiltonian (\ref{BF}) 
we obtain the following set of coupled flow equations \cite{Domanski-01} 
%$\partial_{l} \varepsilon_{\bf k}(l)$, $\partial_{l}E_{\bf q}(l)$ 
%and $\partial_{l}g_{{\bf k},{\bf q}}(l)$ explicitely given 
%equations (16-19) in our previous work \cite{Domanski-01}. 
\begin{eqnarray}
\partial_{l} g_{{\bf k},{\bf q}}(l) &=& - \alpha^{2}_{{\bf k},{\bf q}}
(l) g_{{\bf k},{\bf q}}(l) \\
\partial_{l} \varepsilon_{\bf k}(l) &=& \frac{2}{N} \sum_{\bf q} 
\alpha_{{\bf k},{\bf q}}(l) |g_{{\bf k},{\bf q}}(l)|^{2} n^{(B)}
_{\bf q}\\
\partial_{l} E_{\bf q}(l) &=& \frac{2}{N} \sum_{\bf k}  
\alpha_{{\bf k}-{\bf q},{\bf k}}(l) |g_{{\bf k}-{\bf q},
{\bf k}}(l)|^{2} \nonumber \\ &\times &
\left( - 1 + n^{(F)}_{{\bf k}-{\bf q}
\downarrow} + n^{(F)}_{{\bf k}\uparrow} . \right)
\end{eqnarray}
We have solved them numerically considering 
fermions coupled with bosons on a lattice avoiding thus any need 
for the infrared cutoffs. The fixed point values 
\begin{eqnarray}
\lim_{l\rightarrow\infty}\varepsilon_{\bf k}(l) \equiv  
\tilde{\varepsilon}_{\bf k}, \hspace{1cm} 
\lim_{l\rightarrow\infty}E_{\bf q}(l) \equiv  
\tilde{E}_{\bf q} 
\end{eqnarray}
turned out to reveiled the following features:
\begin{itemize}
\item[{(a)}] for $T\!<\!T_{c}$ the renormalized fermion dispersion 
    $\tilde{\varepsilon}_{\bf k}$  develops a true gap at $\mu$ 
    which evolves into a pseudogap for $T_{c}\!<\!T\!<\!T_{p}$, 
\item[{(b)}] the effective boson dispersion  $\tilde{E}_{\bf q}$ shows 
    the long-wavelength Goldstone mode for $T\!<\!T_{c}$ and its 
    remnants are preserved even in the pseudogap state 
    \cite{Domanski-03}.
\end{itemize}

For a complete information about the fermion and boson spectra 
we need to proceed with transformations for the individual operators  
$\hat{c}_{{\bf k}\sigma}^{(\dagger)}(l)\!\equiv\! e^{\hat{S}(l)}
\hat{c}_{{\bf k}\sigma}^{(\dagger)} e^{-\hat{S}(l)}$ and $\hat{b}_
{\bf q}^{(\dagger)}(l)\!\equiv\! e^{\hat{S}(l)}\hat{b}_{\bf q}^{
(\dagger)} e^{-\hat{S}(l)}$ which is a rather difficult task because
$\hat{S}(l)$ is not known explicitly. Since our primary interest 
is in estimating the particle-hole mixing for the single particle 
fermion excitations we focus on the flow equation 
$\partial_{l}\hat{c}_{{\bf k}\sigma}^{(\dagger)}(l)\!=\!
[\hat{\eta},\hat{c}_{{\bf k}\sigma}^{(\dagger)}(l)]$. 
%The initial derivative 
%\begin{eqnarray}
%$\partial_{l} \hat{c}_{{\bf k}\sigma}(l)_{|_{l=0}} = \frac{\mp 1}
%{\sqrt{N}} \sum_{\bf q}  \alpha_{{\bf q}-{\bf k},{\bf k}}(0)
%\hat{b}_{\bf q} \hat{c}_{{\bf q}-{\bf k},-\sigma}^{\dagger}$
%\label{c_derivative}
%\end{eqnarray}
%  
%(where the sign $-/+$  corresponds to spin $\uparrow/\downarrow$)
%. To satisfy the equation (\ref{c_derivative}) 
%along with its hermitean conjugate 
The generating operator $\hat{\eta}(l)$ chosen according to 
Wegner's prescription \cite{Wegner-94} yields the following 
ansatz for fermion operators \cite{Domanski-03}
\begin{eqnarray}
c_{{\bf k}\uparrow}(l)  &=& 
u_{\bf k}(l) \; c_{{\bf k}\uparrow}  + 
v_{\bf k}(l) \; c_{-{\bf k}\downarrow}^{\dagger}
 \label{c_Ansatz} \\
&+& \frac{1}{\sqrt{N}} \! \sum_{{\bf q} \neq {\bf 0}} \left[
u_{{\bf k},{\bf q}}(l) \; b_{\bf q}^{\dagger}
c_{{\bf q}+{\bf k}\uparrow}  + 
v_{{\bf k},{\bf q}}(l) \; b_{\bf q} 
c_{{\bf q}-{\bf k}\downarrow}^{\dagger}
\right] , \nonumber \\
c_{-{\bf k}\downarrow}^{\dagger}(l) &=& 
- v_{\bf k}^{*}(l) \; c_{{\bf k}\uparrow} +
u_{\bf k}^{*}(l) \; c_{-{\bf k}\downarrow}^{\dagger} 
\label{cbis_Ansatz}\\ & + &
\frac{1}{\sqrt{N}} \! \sum_{{\bf q} \neq {\bf 0}} \left[  
- v_{{\bf k},{\bf q}}^{*}(l) \; b_{\bf q}^{\dagger}
c_{{\bf q}+{\bf k}\uparrow} +
u_{{\bf k},{\bf q}}^{*}(l) \; b_{\bf q}
c_{{\bf q}-{\bf k}\downarrow}^{\dagger}
\right]  ,\nonumber  
\end{eqnarray}
where $u_{\bf k}(0)=1$ and all other coefficients are vanishing
at $l\!=\!0$. The $l$-dependent coefficients must be determined 
from the following set of flow equations \cite{Domanski-03}
\begin{eqnarray}
\partial_{l}  u_{\bf k}(l) & = &  \sqrt{n_{{\bf q}\!=\!{\bf 0}}^{B}} 
\; \alpha_{-{\bf k},{\bf 0}}(l) \; v_{\bf k}(l) 
\label{P_flow} \\ &+& \frac{1}{N} \sum_{{\bf q}\neq{\bf 0}} 
\alpha_{{\bf q}-{\bf k},{\bf q}}(l) \left( 
n_{\bf q}^{B} + n_{{\bf q}-{\bf k}\downarrow}^{F}
\right) v_{{\bf k},{\bf q}}(l), 
\nonumber \\ 
\partial_{l} v_{\bf k}(l) & = & - \;
\sqrt{n_{{\bf q}\!=\!{\bf 0}}^{B}} \; \alpha_{{\bf k},{\bf 0}}(l)
\; u_{\bf k}(l) \label{R_flow} \\ &-& \frac{1}{N} \sum_{{\bf q}\neq{\bf 0}}
\alpha_{{\bf k},{\bf q}}(l) \left(
n_{\bf q}^{B} + n_{{\bf q}+{\bf k}\uparrow}^{F}
\right) u_{{\bf k},{\bf q}}(l), \nonumber \\ 
\partial_{l}  u_{{\bf k},{\bf q}} & = &
\alpha_{-{\bf k},{\bf q}}(l) \; v_{\bf k}(l) ,
\label{p_flow} \\
\partial_{l}  v_{{\bf k},{\bf q}} & = & - \;
\alpha_{{\bf k},{\bf q}}(l)  u_{\bf k}(l) .
\label{r_flow}
\end{eqnarray}
We explored them numerically along with the equations $\partial_{l}
\varepsilon_{\bf k}(l)$, $\partial_{l}E_{\bf q}(l)$, $\partial_{l}
g_{{\bf k},{\bf q}}(l)$ on the 2-dimensional square lattice with 
the initial ($l\!=\!0$) tight-binding dispersion $\varepsilon
_{\bf k}(0)\!=\!-2t \left({\mbox \rm cos}(k_{x}a)+{\mbox \rm cos}
(k_{y}a)\right)$ and the localized boson energy $E_{\bf q}(0)\!=\!E_{0}$.
Moreover, we imposed $g_{{\bf k},{\bf q}}(0)\!=\!g\left(\cos(k_{x}a)
-\cos(k_{y}a)\right)$ to obtain the d-wave symmetry of energy gap 
(pseudogap) below (above) $T_{c}$. We solved the coupled flow 
equations iteratively by the Runge-Kutta method for $E_{0}(0)
\!=\!0.2t$ keeping a fixed charge concentration $n_{tot}=2$ 
when the concentration of fermions $n^{F}\!=\!1+x$ yield the 
realistic value $x\sim 0.1$. In figures \ref{Fig2}-\ref{angle}
we present the results obtained along the antinodal direction
$(0,0) \leftrightarrow (\pi,0)$ i.e.\ for $k_{y}\!=\!0$.

Our ansatz (\ref{c_Ansatz},\ref{cbis_Ansatz}) generalizes the 
standard Bogoliubov-Valatin transformation by including the effect 
of scattering on finite momentum preformed pairs. Influence of such 
scattering shows up in the effective single particle spectral 
function which takes the following form 
\begin{widetext}
\begin{eqnarray}
A({\bf k},\omega)  &=& 
| \tilde{u}_{\bf k}|^{2} \delta \left(\omega\!+\!\mu\!-\!
\tilde{\varepsilon}_{\bf k} \right) 
+ \frac{1}{N} \sum_{{\bf q}\neq{\bf 0}} \left( n_{\bf q}^{B} 
+ n_{{\bf q}+{\bf k}\uparrow}^{F} \right) | \tilde{u}_{{\bf k},
{\bf q}}|^{2}  \delta ( \omega \!+\!\mu\!-\!\tilde{\varepsilon}
_{{\bf q}\!+\!{\bf k}} \!+\! \tilde{E}_{\bf q}) 
\nonumber \\
&+& | \tilde{v}_{\bf k}|^{2} \delta \left( \omega\!-\!\mu\!+\!
\tilde{\varepsilon}_{-{\bf k}} \right) + \frac{1}{N} 
\sum_{{\bf q}\neq{\bf 0}} \left( n_{\bf q}^{B} + n_{{\bf q}-{\bf k}
\downarrow}^{F} \right) | \tilde{v}_{{\bf k},{\bf q}} |^{2} 
\delta ( \omega\!-\!\mu \!+\! \tilde{\varepsilon}_{{\bf q}
\!-\!{\bf k}} \!-\!\tilde{E}_{\bf q}) ,
\label{A_sc}
\end{eqnarray}
\end{widetext}
where $\tilde{u}_{\bf k}$, $\tilde{v}_{\bf k}$ and $\tilde{u}_{{\bf k},
{\bf q}}$, $\tilde{v}_{{\bf k},{\bf q}}$ denote the asymptotic 
$l \rightarrow \infty$ values. We have determined 
them numerically solving the flow equations (\ref{P_flow}-\ref{r_flow})
for the fixed total charge concentration $n_{tot} = 2\sum_{\bf q}
n_{\bf q}^{B} + \sum_{\bf k} \left( n_{{\bf k} \uparrow}^{F} + 
n_{{\bf k} \downarrow}^{F} \right)$.

The structure of spectral function (\ref{A_sc}) indicates that 
besides the narrow peaks (long-lived states) there also forms 
a background of the damped (finite life-time) states. If we neglected
$u_{{\bf k},{\bf q}}$ and $v_{{\bf k},{\bf q}}$ then the flow 
equations (\ref{P_flow},\ref{R_flow}) would simplify to 
$\partial_{l}  u_{\bf k}(l) =  \sqrt{n_{{\bf q}\!=\!{\bf 0}
}^{B}} \; \alpha_{-{\bf k},{\bf 0}}(l) v_{\bf k}(l)$ and 
$\partial_{l} v_{\bf k}(l) = -\sqrt{n_{{\bf q}\!=\!{\bf 0}}^{B}} 
\; \alpha_{{\bf k},{\bf 0}}(l) u_{\bf k}(l)$  yielding the 
invariance $|v_{\bf k}(l)|^{2}+|v_{\bf k}(l)|^{2}\!=\!1$. 
By rewriting the first equation as $\int_{u_{\bf k}(0)=1}
^{u_{\bf k}(\infty)=\tilde{u}_{\bf k}} \frac{d u_{\bf k}(l)}
{\sqrt{1-|u_{\bf k}(l)|^{2}}} = \sqrt{n_{{\bf q}\!=\!{\bf 
0}}^{B}} \; \int_{0}^{\infty} \alpha_{-{\bf k},{\bf 0}}(l) 
dl$ we then right away reproduce the mean-field solution 
$\tilde{u}_{\bf k}, \tilde{v}_{\bf k}=\frac{1}{2} \left(1 \pm 
\frac{\varepsilon_{\bf k}\!-\!\mu}{\sqrt{(\varepsilon_{\bf k}
\!-\!\mu)^{2}+n_{\bf 0}^{B}|g_{{\bf k},{\bf 0}}|^{2}}}\right)$.
In order to go beyond this BCS solution we need to take into 
account the effect of scattering on the finite momentum pairs 
affecting the spectral function (\ref{A_sc}) through the 
coefficients $\tilde{u}_{{\bf k},{\bf q}}$ and $\tilde{v}
_{{\bf k},{\bf q}}$. We shall do it for $T\!>\!T_{c}$.

\section{Particle-hole mixing above {\boldmath $T_{c}$}}

\begin{figure}
\epsfxsize=8cm\centerline{\epsffile{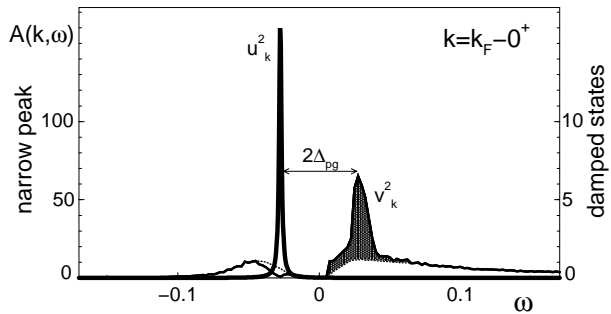}}
\caption{Spectral function $A({\bf k},\omega)$ which consists 
of the long-lived states (we have artificially broadened the 
delta peak into Lorentzian using the units marked on the left 
axis) and the damped fermion states (labels on the right h.s.\ axis) 
slightly below ${\bf k}_{F}$ for $T=0.004D$. The weight of particle
peak $|u_{{\bf k}_{F}}|^{2} \simeq 0.47$ whereas the hole weight 
$|v_{{\bf k}_{F}}|^{2} \simeq 0.19$ (the shaded area) is estimated 
by subtracting the high temperature background.}
\label{Fig3}
\end{figure}

Preformed pairs occupy in the normal phase only the finite 
momenta states (in other words $\langle \hat{b}_{{\bf q}\!=\!
{\bf 0}}\rangle=0$) therefore the equations (\ref{R_flow},
\ref{p_flow}) imply $v_{\bf k}(l)\!=\!0$ and $u_{{\bf k},
{\bf q}}(l)\!=\!0$. The ansatz (\ref{c_Ansatz},\ref{cbis_Ansatz}) 
is thus above $T_{c}$ simplified to
\begin{eqnarray}
\hat{c}_{{\bf k}\uparrow}(l) & = & u_{\bf k}(l) \;
\hat{c}_{{\bf k}\uparrow} \;+ \; \frac{1}{\sqrt{N}} 
\sum_{{\bf q}\neq{\bf 0}} v_{{\bf k},{\bf q}}(l) \; \hat{b}_{\bf q} 
\hat{c}_{{\bf q}-{\bf k}\downarrow}^{\dagger} \label{c_up} \\
\hat{c}_{-{\bf k}\downarrow}^{\dagger}(l) & = &
- \frac{1}{\sqrt{N}} \sum_{{\bf q}\neq{\bf 0}}
v_{{\bf k},{\bf q}}^{*}(l) \; \hat{b}_{\bf q}^{\dagger}
\hat{c}_{{\bf q}+{\bf k}\uparrow} \;+\;
u_{\bf k}^{*}(l) \; \hat{c}_{-{\bf k}\downarrow}^{\dagger} 
\nonumber \\ & &\label{c_down}
\end{eqnarray}
and the corresponding spectral function becomes
\begin{eqnarray}
&&A({\bf k},\omega) = | \tilde{u}_{\bf k}|^{2} 
\delta \left( \omega\!+\!\mu\!-\!\tilde{\varepsilon}_{\bf k} \right) 
\label{spectral} \\&+&\frac{1}{N} 
\sum_{{\bf q}\neq{\bf 0}} \left( n_{\bf q}^{B} + n_{{\bf q}-{\bf k}
\downarrow}^{F} \right) | \tilde{v}_{{\bf k},{\bf q}} |^{2} 
\delta ( \omega\!-\!\mu \!+\! \tilde{\varepsilon}_{{\bf q}
\!-\!{\bf k}} \!-\!\tilde{E}_{\bf q}) .\nonumber
\end{eqnarray}
The first term in (\ref{spectral}) represents the long-lived 
states at the renormalized energies $\tilde{\varepsilon}_{\bf 
k}\!-\!\mu$ whose spectral weight is $|\tilde{u}_{\bf k}|^{2}<1$. 
Remaining part of the spectrum is distributed among the damped 
fermion states. Most of them are almost insensitive to temperature 
and can be regarded as an incoherent background. However, there is 
a certain fraction (very important to us) of a different character 
-- these states emerge around $\omega\!=\!-(\tilde{\varepsilon}
_{\bf k}\!-\!\mu)$ near the Fermi surface as shown in figure 
\ref{Fig2}. Such partly broadened excitation branch, being 
sort of a mirror reflection of the quasiparticle dispersion 
$\tilde{\varepsilon}_{\bf k}\!-\!\mu$, corresponds to the hole 
(particle) contribution for momenta below (above) ${\bf k}_{F}$. 
These ingredients allow us to estimate the Bogoliubov angle 
(\ref{BA}) in the pseudogap state and our procedure for 
determining the particle and hole weights is 
illustrated in figure \ref{Fig3}. 

%    ----------   Figure   ----------    %
%     -------------------------------     %     

To support this treatment we recall some analytical argument 
explaining appearance of the Bogoliubov shadow branch upon approaching 
$T_{c}$ from above. For a decreasing temperature the preformed 
pairs start populating the lower and lower energies so that 
$n_{\bf q}^{B}$ is dominated by the states located just above 
$E_{{\bf q}\!=\! {\bf 0}}$. The resulting spectral function 
(\ref{spectral}) reduces then to 
\begin{eqnarray}
A({\bf k},\omega) & \simeq & | \tilde{u}_{\bf k}|^{2} \delta 
\left( \omega\!+\!\mu\!-\!\tilde{\varepsilon}_{\bf k} \right) 
+ | v_{\bf k}|^{2} \; \frac{\Gamma_{\bf k}/\pi} {(\omega -\!\mu\!
+\!\tilde{\varepsilon}_{{\bf k}})^{2}
\Gamma_{\bf k}^{2}} \nonumber \\ &+&
A_{inc}({\bf k},\omega)
\label{lorentzian}
\end{eqnarray}
where the last term describes solely the structureless 
incoherent background. We obtained $|v_{\bf k}|^{2} $ by 
integrating the spectral function with respect to $\omega$ 
for a given $T$ and subtracting from it the integrated spectral 
function for high temperatures.

%    ----------   Figure 4  ----------    %
\begin{figure}
\epsfxsize=9cm\centerline{\epsffile{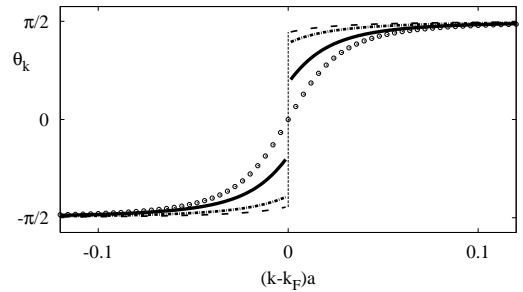}}
\caption{Variation of the Bogoliubov angle estimated in the pseudogap 
state for temperatures $T\!=\!0.004$ (the solid line), $0.007$ (the 
short-dashed curve) and $0.012$ (the long-dashed line). For comparison
we plot by open circles the characteristics of superconducting state
for $T\!=\!0$.}
\label{angle}
\end{figure}
%     -------------------------------     %     

We notice that near ${\bf k}_{F}$ the long-lived state and its 
mirror reflection (a shadow) do not merge because of a finite 
value of the pseudogap $\Delta_{pg}$. 
%Obviously the coefficient 
%$v_{\bf k}^{2}$ is sensitive to temperature (see figure \ref{angle}) 
%but still approaching $k_{F}$ the particle and hole weights 
%varry in a different way. 
%
Figure \ref{angle} shows the calculated Bogoliubov angle as 
a function of momentum measured with respect to the Fermi surface. 
In the pseudogap region the shadow branch has a substantial effect 
on the Bogoliubov angle leading to the p-h mixing near ${\bf k}_{F}$. 
Yet, exactly at the Fermi surface the Bogoliubov angle is discontinuous. 
The BCS-type behavior is finally recovered at temperatures $T\!\leq
\!T_{c}$ as marked by the open circles in figure \ref{angle}. Since 
the magnitude of superconducting gap does not much change 
\cite{Kanigel-07} the Bogoliubov angle is below $T_{c}$
practically frozen (temperature-independent).

\section{Concluding remarks}

We have analyzed the effect of strong superconducting fluctuations 
above the transition temperature \cite{Schmid-70} where the single 
fermions coexist and interact with the preformed pairs. Influence 
of pairs on the single particle excitation spectrum has been studied 
within the selfconsistent RG-like method \cite{Wegner-94}. We have 
found that near $k_{F}$ the renormalized dispersion $\tilde{
\varepsilon}_{\bf k}$ is depleted above $T_{c}$ and additionally 
there appears a shadow branch in the fermion spectrum responsible 
for the particle-hole mixing. We have estimated the particle and 
hole spectral weights thereby determining the Bogoliubov angle
%To our knowledge this has not yet been done above $T_{c}$ 
for the normal state with preformed pairs.
%within any microscopic model.

We have found that momentum dependence of the Bogoliubov angle 
in the pseudogap regime differs qualitatively from its behavior 
for the normal and superconducting states. In the normal state 
(where particle-hole mixing is absent) $\theta_{\bf k}$ changes 
abruptly at ${\bf k}_{F}$ from $-\pi/2$ to $\pi/2$ whereas in 
the superconducting state (below $T_{c}$) the Bogoliubov angle 
continuously evolves between these extreme values over an energy 
regime $\sim \Delta_{sc}$, so that particle and hole excitations 
are mixed with one another. In the pseudogap regime we find that 
$|\theta_{\bf k}|\!\neq\!\pi/2$ but still at the Fermi surface 
the Bogoliubov angle is discontinuous. We hope that STM and ARPES 
techniques would be able to detect such unconventional relation 
between the particle and hole weights predicted for the systems 
with strong pairing fluctuations.

Author acknowledges valuable discussions with J.\ Ranninger,
R.\ Micnas and F.\ Wegner. This work is partly supported by the 
Ministry of Science and Education under the grant NN202187833.

\end{document}